# Title: Broadband nonreciprocal linear acoustics through nonlocal active media

**Authors:** Aritra Sasmal[1], Nathan Geib[1], Karl Grosh[1]†*

**Affiliations:**

[1]Department of Mechanical Engineering, University of Michigan.

†Department of Biomedical Engineering, University of Michigan.

*Correspondence to: grosh@umich.edu.

**Abstract**

The ability to create linear systems that manifest broadband nonreciprocal wave propagation would provide for exquisite control over acoustic signals for electronic filtering in communication and noise control. Acoustic nonreciprocity has predominately been achieved by approaches that introduce nonlinear interaction, mean-flow biasing, smart skins, and spatio-temporal parametric modulation into the system. Each approach suffers from at least one of the following drawbacks: the introduction of modulation tones, narrow band filtering, and the interruption of mean flow in fluid acoustics. We now show that an acoustic media that is nonlocal and active provides a new means to break reciprocity in a linear fashion without these deleterious effects. We realize this media using a distributed network of interlaced subwavelength sensor-actuator pairs with unidirectional signal transport. We exploit this new design space to create media with non-even dispersion relations and highly nonreciprocal behavior over a broad range of frequencies.

**MAIN TEXT**

**Introduction**

Reciprocity in wave-bearing acoustic media is remarkably robust, especially in linear systems, maintained in viscoelastic solids [1], fluid-structure systems [2], and structural-piezoelectric-electrical coupled systems [3]. Further, as is well-established, anisotropy and inhomogeneity, while generating interesting wave propagation phenomenon, do not engender linear nonreciprocity [1]. Acoustic reciprocity, formally introduced by Helmholtz in 1860 (as discussed in [4]) and later generalized by Lyamshev [5] to include fluid-structure interaction and multiple scatters, dictates that the response to a disturbance is invariant upon interchange of the source and receiver. Fluid and solid acoustic media that break reciprocity over broad frequency ranges would enable new and unexplored forms of control over vibrational and acoustic signals, with enormous implications for spectral filtering and duplexing in the communications industry [6], and noise control [7], [8]. Efforts aimed at achieving nonreciprocity in both linear and nonlinear electromagnetic systems have been particularly successful primarily because of the effectiveness of a biasing magnetic field in devices such as the Faraday isolator [9]. These successes have spurred research in analogous acoustic systems where instead of an external magnetic field, introduction of mean flow in the acoustic medium [10] has been used to achieve a high level, narrowband nonreciprocity. Similarly, biasing in a solid using a DC electric field can result in asymmetric damping and nonreciprocal wave propagation in piezoelectric semiconductors [11], [12] as well as in a two-dimensional electron gas coupled to piezoelectric semiconductors [13]. In magnetoelastic and polar media, a DC magnetic field can lead to nonreciprocal effects, although these nonreciprocal effects are often relatively weak (as discussed in [1], [14]). Other approaches to acoustic nonreciprocity rely on breaking the spatial or temporal symmetry in the governing equations by introducing nonlinear interactions [15], [16] or spatiotemporal modulation of the properties of the medium [17], [18]. Theoretical analysis has shown that spatiotemporal modulation of strongly magnetoelastic materials, like Terfenol, and piezoelectric materials, like PZT, can lead to impressive nonreciprocity, as shown in [19]. Both nonlinearity and spatiotemporal modulation introduce secondary tones that require later demodulation or signal processing to prevent signal corruption. To circumvent the disadvantages associated with background bias or spatiotemporal parametric modulation, other studies have utilized collocated sensor-actuator pairs to modulate the wave propagation in the medium in a linear fashion [20]–[22]. To our knowledge, we are the first to exploit a system with distributed control using non-collocated sensor-actuator pairs to introduce inherent violation of parity and time symmetry, and achieve linear acoustic nonreciprocity.

**Results**

In our approach, we use an asymmetric unit cell consisting of a sensor and actuator pair, separated from one another by a subwavelength distance $d_{ff}$, as shown in Fig. 1A. The pairs are arrayed and interlaced along the length of the waveguide. This arrangement breaks spatial symmetry and creates a preferential direction because information is transmitted nearly instantaneously in a unidirectional fashion from sensor to actuator via a distributed amplifier network, while acoustic disturbances propagate bidirectionally at the much slower group velocity of the waveguide. This nonlocal spatial feed forward (NSFF) concept is similar to the canting of the hair cells and phalangeal processes seen in the mammalian cochlea, a feature hypothesized to play a role in wave amplification and dispersion in the hearing organ [23].

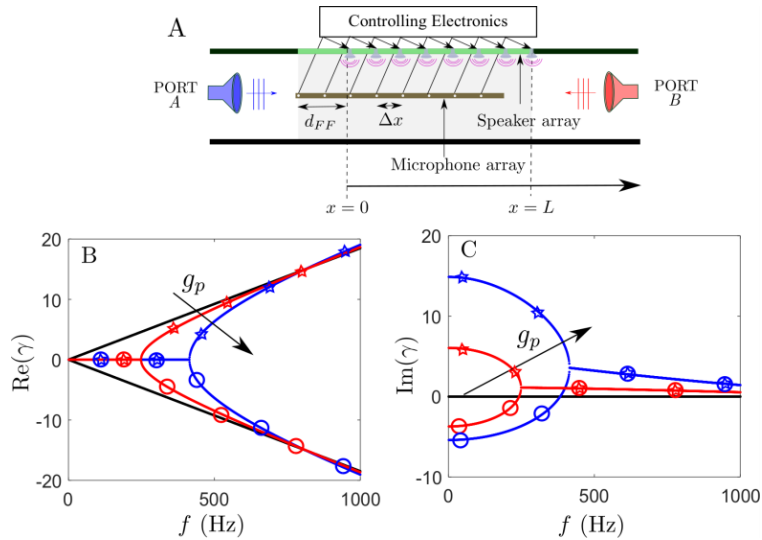

**Fig. 1.** **(A)** Example configuration of the NSFF concept applied to an air-borne acoustic medium. The sensors (microphones) and actuators (speakers) are arrayed along the waveguide and the output of each sensor is fed forward a distance $d_{ff}$ to its corresponding actuator. **(B)** Real part and **(C)** imaginary part of the first two root loci of the complex wavenumber solutions to Eq. (3) for $d_{ff}$=10 cm, and $g_p$ set to three values, 0 m$^{-2}$ (black), 20 m$^{-2}$ (red) and 50 m$^{-2}$ ((blue). Colored stars (circles) are used to delineate the right (left) going waves.

To illustrate this general NSFF concept as a tool to engineer nonreciprocal behavior, we use an airborne acoustic system as shown in Fig. 1A, although this paradigm could be adapted for other wave-bearing media, like piezoelectric or magnetoelastic materials, with appropriate electronic control. First we consider the system in the limit where the acoustic wavelength is much larger than the spacing between successive sensors or actuators ($\Delta x$) so we can treat the active medium as a continuum. The sensed pressure is fed forward to the monopole sources located at a distance $d_{ff}$ downstream. If we assume that the source can be manipulated electronically to precisely match the upstream pressure and that the electronic control is instantaneous, the acoustic source strength can be written as $g_p p(x - d_{ff})$, where $g_p$ is the open loop gain between the sensor and the actuator (see *SI*, Eq. S7). This simplifying assumption will be relaxed later to reflect the dynamics of the acoustic source. With these assumptions, the pressure in the waveguide ($p$) can be modeled using a modified version of the one-dimensional Helmholtz equation with an additional pressure-proportional source term as

$$\frac{d^2 p(x)}{dx^2} + k^2 p(x) = g_p p(x - d_{ff}) \tag{1}$$

where $c$ is the acoustic speed, $\omega$ is the radian frequency (assuming an $e^{-i\omega t}$ time dependence), and $k = \frac{\omega}{c}$. The gain $g_p$ is non-zero for $x \in (0, L)$ and is zero elsewhere. To show the nonreciprocity of the NSFF, let $p^I(x)$ be the solution of Eq. 1 due to a point source $Q^I$ at $x = x_1$ and $p^{II}(x)$ the solution due to a point source $Q^{II}$ at $x = x_2$, where $x_1 < 0$ and $x_2 > L$. Following standard arguments typically used to prove reciprocity in acoustics [1], we find

$$-\int_0^L g_p \left( p^{II}(x) p^I(x - d_{ff}) - p^I(x) p^{II}(x - d_{ff}) \right) dx = p^{II}(x_1) Q^I - p^I(x_2) Q^{II} \tag{2}$$

so that acoustic reciprocity, given by $p^{II}(x_1) Q^I = p^I(x_2) Q^{II}$ [24], holds only at exceptional frequencies when the left-hand-side integral vanishes. The spatial separation of the sensor and the actuator and the unidirectional sensor-signal transmission are the crucial elements in achieving inherent nonreciprocity in the NSFF system. This is fundamentally different from the case where active elements of an acoustic waveguide are coupled via a bidirectional transmission line [25], because such a system is reciprocal. The nonlocal approach is also different from case where the sensor and source are collocated and local impedance modification or bianisotropy is utilized to achieve nonreciprocity [20], [21], because the nonlocality, even though subwavelength, affords addition flexitility in achieving nonreciprocity.

To further investigate the nonreciprocal wave characteristics of the NSFF, we assume harmonic waves of the form $p_0 e^{i\gamma x}$ to obtain the dispersion relation in the active region (Eq. 1) given by

$$-\gamma^2 + k^2 = g_p e^{-i\gamma d_{ff}} \tag{3}$$

where $\gamma$ is the wavenumber. Owing to the exponential term on the right-hand side of Eq. 3 there are an infinite number of complex root loci and the equation is not even in $\gamma$. In order to show the evolution of the complex wavenumber-frequency loci with increasing gain, we plot the real part of the wavenumber in Fig. 1B and the imaginary part in Fig. 1C for the first two root loci. Two nonzero values of $g_p$ are chosen, $g_p = 50 \text{ m}^{-2}$ (in blue) and $g_p = 20 \text{ m}^{-2}$ (in red). The nondispersive and purely real wavenumber loci $\gamma_{+/-} = \pm \omega/c$ of the passive case ($g_p = 0 \text{ m}^{-2}$) are shown for reference with black lines. For both active loci, the allowed waves are purely evanescent at low frequencies and asymmetric about the ordinate, indicating directionally dependent phenomena. The loci for both choices of nonzero gains exhibit a bifurcation point beyond which the solutions exhibit decay in the left-to-right direction ($A \rightarrow B$) and growth in the right-to-left direction ($B \rightarrow A$), demonstrating spatial nonreciprocity of the active waveguide. These gains yielded stable temporal solutions for the unbounded case, as confirmed by simulating the impulse response of the active waveguide. Increasing the gain changes the asymmetry of the evanescent component of the wavenumber, increases the frequency where the bifurcation point occurs, and eventually results in instability.

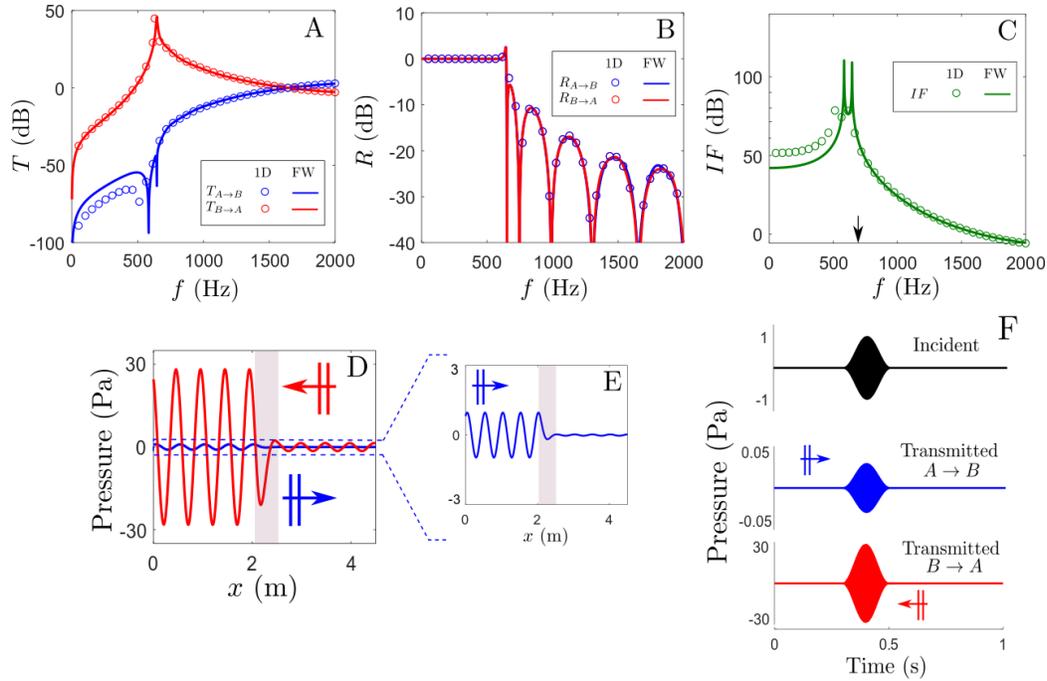

**Fig. 2.** The **(A)** transmission and **(B)** reflection coefficients of the discrete realization of the active waveguide for a wave traveling from port $A$ to $B$ (blue) and from port $B$ to $A$ (red) as obtained from full wave (FW) simulations (solid lines) and 1-D simulations (circles). **(C)** The isolation factor ($IF$) derived from the transmission and reflection spectra from (A) and (B). **(D)** FW simulation of the spatial pressure field for a plane wave incidence from port $A$ (blue) and port $B$ (red) at 692 Hz (frequency shown with black arrow in (C)) showing 29 dB of amplification for propagation from $B$ to $A$ and 31 dB of attenuation for propagation from $A$ to $B$. A magnified view of the wave propagating from $A$ to $B$ is shown in **(E)**. **(F)** The time evolution of the wave envelopes of the transmitted pressure at the output of the waveguide due to a 0.2 s cosine squared pulse centered at 692 Hz incident from port $A$ (blue) and port $B$ (red) are shown. The incident pulse is shown (black) for reference. Notice the different pressure scales associated with the incidence directions in Figs. 3D-F. A constant and uniform gain of $g_d$=4.5 m$^{-1}$ has been used for each sensor-actuator pair in all simulations.

To determine if the nonreciprocity seen in the continuous system is conveyed to a system composed of discrete sensors and actuators, we consider an array of $N = 10$ uniformly spaced pairs ($\Delta x = 5$ cm) arranged in the active section in an infinite acoustic duct as shown in Fig. 1A. We retain the assumption that the electronics can provide the gain necessary to guarantee that the acoustic source strength of each actuator is equal to the discrete gain, $g_d$, times the measured pressure at a distance $d_{ff} = 10$ cm upstream, similar to the source term in Eq. 1. We modeled this numerically in two ways. First, we used one-dimensional (1D) acoustic theory, with the actuators idealized as point sources. Second, we used a full-wave (FW) solution that consisted of a complete three-dimensional finite element acoustic model in Comsol Multiphysics that included the finite extent of the sources, treated as boundary velocity forcing, and three dimensionality of the fluid domain. Parameters for the 1D and FW models are given in the *SI*. We define the transmission coefficient ($T$) as the ratio of the amplitude of the transmitted and the incident pressure field, and the reflection coefficient ($R$) as the ratio of the amplitude of the reflected and the incident field, expressed in dB. For a plane wave incident from port $A$, the subscript $A \to B$ is used while the subscript $B \to A$ represents the opposite situation. As shown in Fig. 2A, $T_{A \to B} \neq T_{B \to A}$ over the frequency range plotted (except at an exceptional frequency) resulting in a non-symmetric scattering matrix, demonstrating the nonreciprocal nature of the system. The reflection coefficients (Fig. 2B) are equal in amplitude, $|R_{A \to B}| = |R_{B \to A}|$ but differ in phase by $2kd_{ff}$ radians (see proof in the *SI*) for our equispaced sensor-actuator system. This is in stark contrast with $\mathcal{PT}$ symmetric systems, where the transmission coefficients from either direction are the same and the reflection coefficients differ [26]. Further, if the actuator has sufficient authority to deliver pressure at very low frequencies, this system reflects incoming waves from both directions at those frequencies, acting as a subwavelength wall for sound. The FW simulations are in good agreement with the 1D acoustic theory in this frequency range. The degree of nonreciprocity quantified by the isolation factor (*IF*), defined as the difference of $T_{B \to A}$ and $T_{A \to B}$, exceeds 40 dB over a broad range of frequencies from DC up to 800 Hz, as shown in Fig. 2C, and displays a 20 dB *IF* bandwidth of more than 1 kHz.

To further elucidate the effectiveness of the NSFF, the spatial variation of the real part of the total pressure field due to incidence of 692 Hz plane wave from port *A* is shown in Fig. 2D and incidence from port *B* in Fig. 2E. This frequency was chosen to establish the efficacy of the active waveguide away from the maximum *IF*. The plane wave incident from port *B* (Fig. 2D) is amplified by 29 dB whereas the wave incident from port *A* (Fig. 2E) is attenuated by 31 dB, leading to a remarkable net acoustic *IF* of 60 dB. To determine the effectiveness of the distributed active media under transient loading, we simulated the response of the active waveguide to a cosine squared windowed incident pulse 0.2 ms in duration and centered at frequency of 692 Hz, the envelope of which is shown in Fig. 2F. Time domain calculations show that the transmitted wave packets exhibit minimal distortion, and the wave packet traversing from port *B* to *A* (red line) is amplified, whereas the transmitted wave packet traveling from port *A* to *B* (blue line) is reduced, consistent with the 60 dB *IF* predicted by the steady state response. The system was shown to be stable by casting the solution of the 1D model into the canonical closed loop transfer function form and applying the Nyquist stability criterion as outlined in the *SI*. FW solution stability was confirmed by finding the impulse response, consistent with the transient results shown in Fig. 2F which also show stability.

To verify the viability of the spatial feed-forward control with real electromechanical transducers, we relaxed the assumption that the source strength is precisely equal to the sensed pressure, as introduced in Eq. 1. Instead, we used the voltage output from each microphone (sensor) multiplied by a gain factor $g_d$ as the input voltage to the corresponding electrodynamic speaker (actuator) to simulate a real experiment. Using standard electrodynamic driver theory [24], we modeled each of the 10 sources with the nominal Thiele-Small parameters for a typical minispeaker, as documented in the *SI*. Fig. 3A shows the *IF* spectrum for the maximum stable discrete gain, $g_d^{max}$, $0.5 g_d^{max}$, $0.1 g_d^{max}$ and passive waveguide ($g_d = 0$ m$^{-1}$) to show the change in the *IF* spectrum with decreasing gain. Fig. 3B shows the spatial variation of the pressure field at 900 Hz corresponding to *IF* of 50 dB. For a 1 Pa incident field, the voltage applied to the speakers remained under the maximum voltage rating for this speaker over the entire range of frequencies. Our calculations predict a maximum stable *IF* of 50 dB at 886 Hz. We define $\Delta f_{IF}$ as the 20 dB *IF* bandwidth, and calculated it to be 456 Hz for this system, equal to 51% of the peak *IF* frequency. Other studies utilizing linear mechanisms to achieve nonreciprocity have reported peak *IF*s of around 40 dB ($\Delta f_{IF}$=4 Hz) for the acoustic circulator [15] and 25 dB ($\Delta f_{IF}$=250 Hz) for the Willis metamaterial [21]. Hence, this proposed mechanism has the potential to exceed the maximum level and bandwidth achieved by other approaches [15], [21] without disrupting mean fluid flow. Further, the *IF* spectrum can be easily manipulated by electronically modulating $g_d$, providing a highly flexible mechanism for *in situ* optimization of the NSFF system for specific applications.

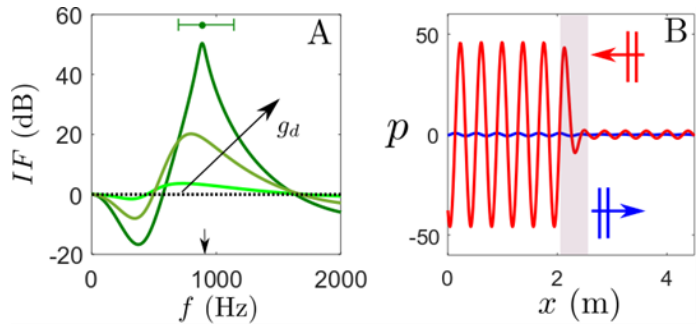

**Fig. 3.** Nonreciprocity in the NSFF system with actuators modeled as electrodynamical speakers. **(A)** The curves show the transition of the IF for $g_d = 0$ (black) to the highest stable gain $g_d^{max}$ (light to dark shades of green). The 20 dB IF bandwidth is shown as a horizontal line at the top of the plot with the peak IF frequency indicated with a filled circle. **(B)** FW simulation of the spatial distribution of the pressure field for a 900 Hz unit amplitude plane wave propagating from $B$ to $A$ (red) and from $A$ to $B$ (blue). The active section is shown with the shaded box.

**Discussion**

We have shown that it is possible to induce linear broadband nonreciprocity in acoustic systems, essentially creating a new stable media using the NSFF mechanism. This mechanism consists of an array of interlaced subwavelength sensor-actuator unit cells (the total active region can be sub- or supra-wavelength). Although we have demonstrated the approach using a fluid-acoustic medium, this technique can be adopted and applied to many different wave-bearing media and systems. For instance, the locally sensed force or strain in either an interdigitated surface acoustic wave device [27][28][29] or a layered stack of bulk-wave piezoelectric elements [30][31] can be fed forward to actuator elements using the NSFF approach, creating a preferred direction and nonreciprocity. The NSFF approach expands the design space, holding the potential to enhance the desired capability of the device (e.g., filtering or sound output). An extensively studied prototype for wave propagation and control in dispersive systems is an elastic beam bounded to piezoelectric patches arrayed down the beam. When the piezoelectric elements are electrically interconnected by a transmission line, a coupled elastic-electric waveguide is created [32]. While this coupled waveguide system can be designed to achieve excellent stop-band behavior or high losses, it is still reciprocal. By breaking the bidirectionality of the transmission line using the feed forward distributed control of the NSFF, these reciprocal systems would be converted to nonreciprocal ones. Another popular approach is to use collocated sensor-actuator patch approaches to control wave propagation on beams, as in [33]. These too can be converted to nonreciprocal systems by feeding forward the control signal to the neighboring patch. Finally, one can also envision creating nonreciprocal anisotropy in two-dimensional media, potentially enabling one-way waveguiding. Hence, our theoretical work opens up the possibility of reconfiguring a vast array of well-studied systems rendering them nonreciprocal. While we have used a gain which is spatio-spectrally constant, exploring the vast design space associated with the spatio-spectral variation of the amplitude and phase of the gain associated with each sensor-actuator pair as well as the distance between them holds great potential for noise control as well as for enhancement of the performance of electromechanical filters and amplifiers.

**Acknowledgments**

**General**: We would like to acknowledge Prof. Bogdan Popa for his helpful discussions.

**Funding:** This study was funded by the National Science Foundation grant DCSD-1761300 and the National Institute of Health grant R01 DC04084.

**Author contributions:** Describe the contributions of each author (use initials) to the paper.

**Competing interests:** The authors declare no conflict of interest.

**Data and materials availability:** None

# Supplementary Materials and Methods

## Proof of Nonreciprocity in the Long Wavelength Limit

In the long wavelength regime, where the acoustic wavelength is much larger than the spacing between consecutive actuators ($\lambda \gg \Delta x$), the discrete acoustic sources can be treated as a continuum source of strength $g_p p(x - d_{ff})$. In this limit, the pressure in the waveguide ($p$) can be modeled using the one dimensional Helmholtz equation as

$$\frac{d^2 p(x)}{dx^2} + k^2 p(x) = \begin{cases} g_p p(x - d_{ff}), & (0 < x < L) \\ 0, & \text{otherwise} \end{cases} \tag{S1}$$

where $k = \frac{\omega}{c}$, $\omega$ is the angular frequency, and $c$ is the acoustic speed in air. The active section of the waveguide ($g_p \neq 0$) extends from $x = 0$ to $x = L$. Let $p^I(x)$ be the solution of Eq. S1 due to a point source $Q^I$ at $x = x_1$ and $p^{II}(x)$ the solution due to a point source $Q^{II}$ at $x = x_2$, where $x_1 < 0$ and $x_2 > L$.

$$\frac{d^2 p^I(x)}{dx^2} + k^2 p^I(x) = \begin{cases} g_p p(x - d_{ff}), & (0 < x < L) \\ Q^I \delta(x - x_1), & \text{otherwise} \end{cases} \tag{S2A}$$

$$\frac{d^2 p^{II}(x)}{dx^2} + k^2 p^{II}(x) = \begin{cases} g_p p(x - d_{ff}), & (0 < x < L) \\ Q^{II} \delta(x - x_2), & \text{otherwise} \end{cases} \tag{S2B}$$

Multiplying Eq. S2A by $p^{II}$ and Eq. S2B by $p^I$, subtracting and integrating from $(-\infty, \infty)$ along with the continuity of pressure and velocity yields

$$-\int_0^L g_p \left( p^{II}(x) p^I(x - d_{ff}) - p^I(x) p^{II}(x - d_{ff}) \right) dx = p^{II}(x_1) Q^I - p^I(x_2) Q^{II}. \tag{S3}$$

Acoustic reciprocity requires $p^{II}(x_1) Q^I - p^I(x_2) Q^{II} = 0$, which is possible only if $d_{ff} = 0$. Since the non-local spatial feedforward (NSFF) mechanism requires that the sensors and actuators be non-collocated, i.e. $d_{ff} \neq 0$, this system is inherently nonreciprocal, except at certain exceptional frequencies.

# Modeling and Stability Analysis of the Discrete NSFF System

Stability in active feedback systems is of paramount interest while designing the system. We studied the stability of the NSFF system to delineate our operational bounds for the gain $g_d$ between the sensors and actuators. We demonstrate our calculations by considering the air-borne acoustic feedforward system is shown in Fig. S1. The N sensors and actuators are aligned along the duct and the spacing between two consecutive sensors is $\Delta x$. The Green's function for a one dimensional infinite acoustic duct is given by

$$G(x,y) = \frac{1}{2\gamma} e^{i\gamma|x-y|},$$

where $\gamma = \frac{\omega}{c}$ is the acoustic wavenumber. The pressure at each sensor can be written as a superposition of the pressure from the incoming plane wave and the pressures due to each of the N

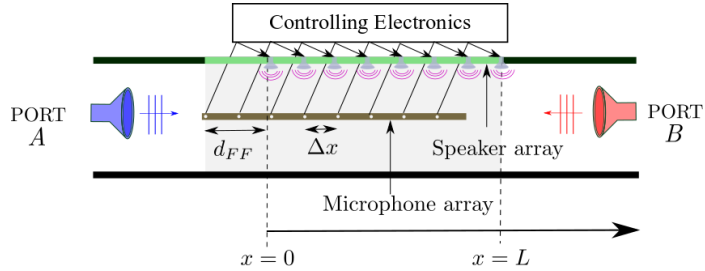

**Fig. S1.** Schematic of the NSFF system. The array of probes and actuators are arrayed along the duct, offset from each other by a distance $d_{ff}$. The distance between consecutive actuators (or probes) is equal to $\Delta x$. The pressure measured by each probe is electronically modulated and fed to the actuator at a distance $d_{ff}$ upstream. The duct is assumed to be infinite in both directions and the waves entering from either ports are assumed to be plane waves.

actuators as

$$P_p^i = p_0 e^{\pm i\gamma x_i} + \frac{1}{2i\gamma} \sum_{j=1}^{N} e^{i\gamma|x_{pi}-x_{sj}|} S(x_j), \qquad (S4)$$

where the pressures at the $i^{th}$ sensor located at position $x_{pi}$ is $P_p^i$ and the strength of the acoustic source due to the $j^{th}$ actuator located at position $x_{sj}$ is $S(x_j)$. If the source strength of each actuator ($S_j$) is set to the value of the pressure detected by its corresponding upstream sensor ($p_j$) modulated by a gain $g_d$, the pressures at the $N$ sensors can be written in a vector form as

$$\boldsymbol{P}_p = p_0 (I - g_d \boldsymbol{G})^{-1} e^{\pm i\gamma \boldsymbol{X}_p}, \qquad (S5)$$

where $\boldsymbol{P}_p = [P_p^1, P_p^2 \dots P_p^N]^T$, $\boldsymbol{X}_m = [x_{p1}, x_{p2}, \dots x_{pN}]^T$, and $\boldsymbol{G}_{ij} = \frac{1}{2i\gamma} e^{i\gamma|x_{pi}-x_{sj}|}$. The above equation is in the canonical form of a multiple input multiple output closed loop control system, and we determined the range by determining the winding number of the scalar function $\det(I - g_d \boldsymbol{G})$ along the Nyquist contour. Fig. S2 shows the variation of winding number with gain for 10 sensor-actuator pairs as discussed in the main text. The gains corresponding to trivial winding numbers are associated with stable NSFF system and are shown with the shaded region.

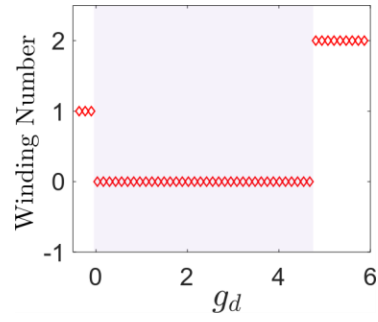

**Fig. S2.** The winding number variation with gain for N=10 pairs of probes and actuators. The source strength is assumed to be equal to the pressure at the upstream probe modulated by a scalar gain $g_d$. The system is stable when the winding number of $\det(I - g_d G)$ along the Nyquist contour is equal to zero. Using this design, the system is stable when the discrete gain $g_d \in (0, 4.6)$, and the stability boundary is shown with the shaded box.

## NSFF Using Electrodynamical Speakers

We used the Thiele Small parameters for the 1-1/4 inch CE30P-4 Dayton mini speakers to model the NSFF system discussed in the main text. The parameters are tabulated in Tab. S1. The transfer function of the velocity of the speaker membrane ($V_{sp}$) due to voltage applied to the terminals ($\phi$) is given by

$$\frac{V_{sp}}{\phi} = \frac{BL}{Z_e(Z_m + (BL)^2/Z_e)},$$

where $Z_e(s) = Ls + R$, $Z_m = ms + \frac{k}{s} + c$ and $s = -i\omega$.

Tab. S1: Parameters associated with the model with microphones and speakers,

| Parameter | Description | Value in SI |
|---|---|---|
| BL | Magnetic coupling | 1.906 Wb/m |
| $\rho$ | Density of air | 1.2 Kg/m$^3$ |
| R | Electrical resistance | 3.508 $\Omega$ |
| L | Electrical self-inductance | 1.64 E-4 H |
| m | Mass of membrane | 6.08E-4 Kg |
| k | Stiffness of membrane | 4651.2 N/m |
| c | Damping of membrane | 0.486 Kg/s |
| $\Gamma$ | Sensitivity of microphone | 0.5 V/Pa |
| $S_{duct}$ | Duct cross-section | 6.4516E-4 m$^2$ |
| $S_{sp}$ | Actuator diaphragm cross-section | 3.1416E-4 m$^2$ |

The output of the microphone (of sensitivity $\Gamma = \frac{\phi}{p}$) was amplified through a gain $g_d$ using an acoustic amplifier and supplied to the input terminals of the speaker. The complete transfer function between the velocity of the speaker $V_{sp}$ and the pressure at its corresponding microphone (located at a distance $d_{ff}$ upstream) is given by

$$V_{sp} = G_{sp}p(x - d_{ff}) = \Gamma\, g_d \frac{BL}{Z_e(Z_m+(BL)^2/Z_e)} p(x - d_{ff}). \tag{S6}$$

Using Eq. S6, the acoustic source strength $S(x_j)$ in Eq. S4 can be written in terms of the pressure at the j$^{th}$ microphone $P_{pj}$ as

$$S(x_j) = \left[-i\omega\rho V_{sp} \frac{S_{sp}}{S_{duct}}\right] = G_{sp}P_{pj}, \tag{S7}$$

where $G_{sp} = \Gamma g_d \frac{-i\omega S_{sp}}{S_{duct}} \frac{BL}{Z_e(Z_m+(BL)^2/Z_e)}$, $S_{duct}$ is the cross-sectional area of the duct, and $S_{sp}$ is the area of the speaker diaphragm. Stability calculations similar to Fig. S2 yields a stable discrete gain boundary of $g_d \in (-0.04, 0.088)$.

**Full Wave (FW) Simulations**

We performed 3D full wave simulations in Comsol Multiphysics v5.4 using the pressure acoustics module. The schematic of the modeled system is shown in Fig. S3. We imposed radiation boundary conditions at either end (shaded green) to model the infinite waveguide. The sensors were positioned along the center of the duct, at a distance $d_{ff}$ upstream from its actuator pair (blue circles). A representative sensor-actuator pair is shown in Fig. S3 where the pressure at the cross-section shaded red is fed forward a distance $d_{ff}$ to its actuator (blue circle) through a gain $G_{sp}$. For the FW model of the system, the actuators are modeled as boundary velocity forcing to replicate the diaphragm of electrodynamical speakers operating in the piston-mode with an average velocity $V_{sp}$ given by Eq. S6 and the nominal values for microphone and electrodynamical speakers (Tab. S1) were used calculate the acoustic response. For the ideal actuator assumption (source strength equal to the upstream pressure, $S(x_j) = g_d P_{pj}$), the velocity of the j$^{th}$ actuator was set to be

$$V_{sp} = g_d \frac{S_{duct}}{-i\omega\rho S_{sp}} P_{pj},$$

where $P_{pj}$ is the pressure at the corresponding upstream probe.

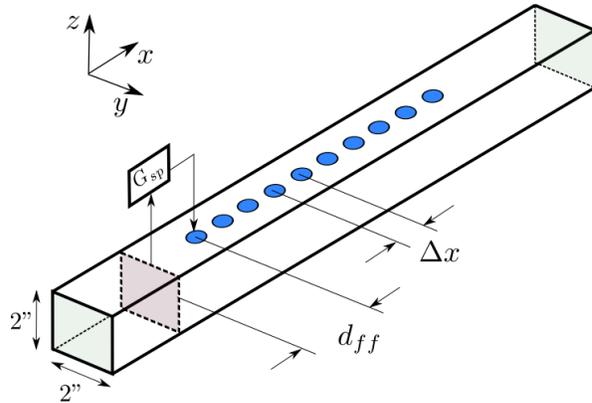

**Fig. S3.** The schematic of the 3D waveguide as modeled in finite element. The acoustic source strength of the actuator is controlled by the signal from a probe placed at a distance $d_{ff}$ upstream modulated by the gain $G_{sp}$.

# Equality of Reflection Coefficients in the NSFF System

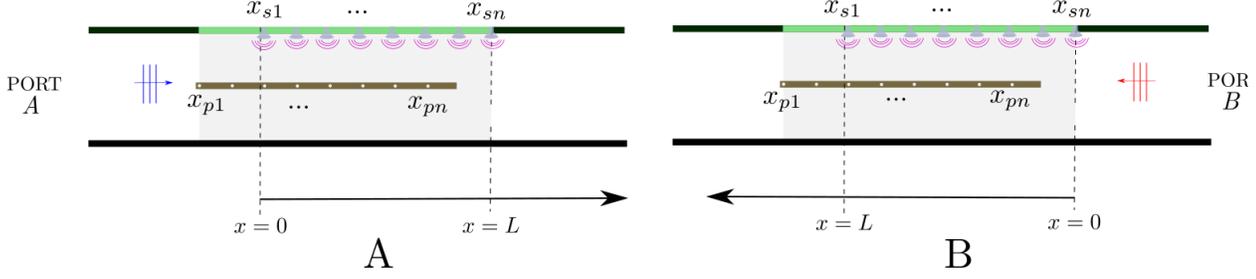

**Fig. S4.** Calculation of transmission and reflection coefficients for the NSFF mechanism using 1D acoustic theory for waves incident from the **(A)** left and from the **(B)** right.

Let the coordinates of the sensors be at $X_p = \{x_{p1}, x_{p2}, ... x_{pN}\}^T$ and the coordinates of the actuators be at $X_s = \{x_{s1}, x_{s2}, ... x_{sN}\}^T$ as shown in Fig. S4A. Without loss of generality, let $x_{s1} = 0$ and $x_{sN} = L$. The coordinates of the $m^{th}$ sensor can be written as $x_{pm} = (m-1)\Delta x - d_{ff}$, and the coordinate of the $n^{th}$ actuator can be written as $x_{sn} = (n-1)\Delta x$. The total pressure field at any location $x$ can be written as a linear superposition of the incident pressure field, $p_{inc} = p_o e^{\pm i\gamma x}$ and the scattered pressure field, $p_{sc} = \frac{g_d}{2i\gamma} \sum_{j=1}^{N} e^{i\gamma|x_{sj}-x|} P_{pj}$, where $P_{pj}$ is the pressure detected at the $j^{th}$ sensor and are given by Eq. S4. The reflection and transmission coeffcients can be written as

$$R_{A \to B} = \sum_{n=1}^{N} \sum_{m=1}^{N} \frac{1}{2i\gamma} e^{i\gamma(n-1)\Delta x} H_{nm} e^{i\gamma(m-1)\Delta x} e^{-i\gamma d_{ff}},$$
$$T_{A \to B} = e^{i\gamma(N-1)\Delta x} + \sum_{n=1}^{N} \sum_{m=1}^{N} \frac{1}{2i\gamma} e^{i\gamma(n-1)\Delta x} H_{nm} e^{i\gamma(m-1)\Delta x} e^{-i\gamma d_{ff}}, \quad (S8)$$

where $\gamma$ is the acoustic wavenumber and $H = g_d(I - gG)^{-1}$. Here $G_{ij}$ is the Green's function from the $j^{th}$ actuator to the $i^{th}$ sensor.

To simplify the calculation of the reflection and transmission coefficients for waves impinging from port B, we choose a different set of coordinates such that $x_{s1} = L$ and $x_{sn} = 0$, as shown in Fig. S4B. In this coordinate system, the coordinate of the $m^{th}$ sensor can be written as $(N-m)\Delta x + d_{ff}$, and the coordinate of the $n^{th}$ actuator is $(N-n)\Delta x$. Note that the choice of coordinates does not affect the calculation of the reflection and transmission coefficients. The reflection and transmission coefficents for a wave impinging on the NSFF system from port B can be written as

$$R_{B \to A} = \sum_{n=1}^{N} \sum_{m=1}^{N} \frac{1}{2i\gamma} e^{i\gamma(N-n)\Delta x} H_{nm} e^{i\gamma(N-m)\Delta x} e^{ikd_{ff}},$$
$$T_{B \to A} = e^{i\gamma(N-1)\Delta x} + \sum_{n=1}^{N} \sum_{m=1}^{N} \frac{1}{2i\gamma} e^{i\gamma(N-n)\Delta x} H_{nm} e^{i\gamma(N-m)\Delta x} e^{i\gamma d_{ff}},$$

Now, under the assumption that $d_{ff}$ and the sensor-actuator gain $g_d$ is constant across all pairs, the matrix $G$ is a persymmetric matrix, i.e. $G_{m,n} = G_{N-n+1,N-m+1}$. This results in $H = g_d(I - g_d G)^{-1}$ inheriting persymmetry, and consequently $H_{m,n} = H_{N-n+1,N-m+1}$. The reflection and transmission coefficients can now be simplified to

$$R_{B \to A} = \sum_{n=1}^{N} \sum_{m=1}^{N} \frac{1}{2i\gamma} e^{i\gamma(n-1)\Delta x} H_{nm} e^{i\gamma(m-1)\Delta x} e^{i\gamma d_{ff}},$$

$$T_{B \to A} = e^{i\gamma(N-1)\Delta x} + \sum_{n=1}^{N} \sum_{m=1}^{N} \frac{1}{2i\gamma} e^{i\gamma(n-1)\Delta x} H_{nm} e^{ik(m-1)\Delta x} e^{i\gamma d_{ff}}, \tag{S9}$$

Comparing Eq. S8 and Eq. S9, we see that $|R_{A \to B}| = |R_{B \to A}|$ and the reflection coefficients differ in phase by $2\gamma d_{ff}$ radians. The transmission coefficients do not have any proportional relationship.

It is remarkable that the persymmetry in the NSFF system yields very different results from $\mathcal{PT}$ symmetric reciprocal systems. In $\mathcal{PT}$ symmetric systems, calculation of reflection and transmission coefficients yields $|T_{A \to B}| = |T_{B \to A}|$ (which support reciprocity) and $|R_{A \to B}| \neq |R_{B \to A}|$, whereas in the NSFF system, $|T_{A \to B}| \neq |T_{B \to A}|$ and $|R_{A \to B}| = |R_{B \to A}|$.